\documentclass[11pt]{article}
\usepackage[dvips,letterpaper,margin=1.0in,bottom=0.5in, top=0.5in]{geometry} 
\usepackage[english]{babel}
\usepackage{epstopdf}
\usepackage{pdfpages}
\usepackage{float}
\usepackage{enumerate}

\usepackage{amsmath} 
\usepackage{dsfont}
\usepackage{cancel}
\usepackage{amsthm} 
\usepackage{amssymb}	
\usepackage{amsmath}
\usepackage{graphicx} 
\usepackage{empheq}
\usepackage[absolute,overlay]{textpos}
\usepackage{xcolor}
\usepackage{hyperref}
\usepackage{tikz}
\usepackage{relsize}

\begin{document}

\title{Universality and Invariance in Hegselmann-Krause Opinion Dynamics: Proof of Three Conjectures}

\date{\today}

\author{Paolo Molignini, AlbaNova University Center, Stockholm University}

\maketitle

\begin{abstract}
Three conjectures from Ref.\cite{Hegselmann:2023} about the Hegselmann-Krause opinion dynamics and the structure of $\epsilon$-switches are proved.
The first conjecture states that the number of $\epsilon$-switches for any given initial opinion distribution is always finite, guaranteeing that the algorithm for enumerating them terminates.
The second conjecture concerns the relationship between the dynamics of two consecutive $\epsilon$-switches, showing that the opinion evolution is identical up to the switch time.
The third conjecture establishes the invariance of the dynamics under positive-affine transformations of the initial distribution, with a corresponding rescaling of all $\epsilon$-switch values.
Together, these results provide a formal foundation for the empirical observations reported in Ref.\cite{Hegselmann:2023} and offer a step towards a systematic classification of BC-processes based on their initial conditions.
\end{abstract}

\section{Introduction}
Opinion dynamics is a field at the intersection of physics, mathematics, and the social sciences that studies how the distribution of individual opinions in a population changes over time as a result of interpersonal interactions~\cite{Fortunato:2004, Castellano:2009, Galam:2016, Sirbu:2017}. 
Models in this area aim to capture qualitative features such as consensus formation, polarization, clustering into subgroups, or persistent disagreement. 
Despite their simplicity, such models often display rich nonlinear phenomena and are widely used to explore the mechanisms that drive social influence, the spread of ideas, and collective decision-making.

Among the best-known frameworks in this field is the Hegselmann-Krause (HK) model~\cite{Hegselmann-Krause:2002}. 
This model describes a set of agents, each holding a real-valued opinion, who update their opinions synchronously by averaging those of other agents that are sufficiently close to them in opinion space. 
This bounded confidence dynamics is nonlinear and mimics the behavior observed in real-world social systems where individuals interact selectively, such as political discourse, product adoption, or cultural assimilation.
The HK model is minimalistic yet capable of producing complex dynamics, such as the emergence of opinion clusters, non-monotonic transitions in the number of clusters as the confidence bound varies, and long-lived transient states.

Recently, the HK model has been revisited with the tools of fractional arithmetic~\cite{Hegselmann:2023}. 
This different approach was accompanied by the introduction of novel methods to study the dynamics of the bounded confidence model, most notably $\epsilon$-switches.
This new concept refers to an increasing sequence of $\epsilon$ values that trigger changes to the dynamics given the same initial configuration of agents' opinions.
With the combination of fractional arithmetic and $\epsilon$-switches, it is possible to set up an algorithm that exactly maps out all the BC-processes for a given initial distribution~\cite{Hegselmann:2023}.
This procedure can be used to also map out all possible non-monotonicities that arise in the steady-state distribution as $\epsilon$ is varied and -- in a nutshell -- completely describe the behavior of the dynamics given the initial state and the confidence level.

During the expostulation of this algorithm, Ref.~\cite{Hegselmann:2023} states three conjectures that are supported by strong numerical evidence, but are not formally proven.
They pertain to the size of the set of $\epsilon$-switches, to the relationship between the dynamics of consecutive switches, and to the operations that leave the set of $\epsilon$-switches (and the underlying dynamics) invariant.
The aim of this work is to provide a concrete proof for each of the three subjects above, which formally confirms the empirical observations encountered in the numerics.

\section{Definitions}

Before stating the three conjectures explicitly, we provide a summary of the notation and terminology used in this work, which largely repeats the notation used in Ref.~\cite{Hegselmann:2023}.

The Hegselmann-Krause model describes a set $\mathcal{A}$ of $n$ agents, labelled with integers from $1$ to $n$.
Each agent $i$ starts off at time $t=0$ with a certain opinion value $x_i(0) \in [0,1]$.
Time is discretized into integers $t=0, 1, 2, \dots$ and the opinions evolve from one time step to the next according to an update rule, which will define below.
The opinion of agent $i$ at time $t$ is denoted as $x_i(t) \in [0,1]$.
For simplicity, we can compactly store all the opinions into a vector
\begin{equation}
X(t) \equiv (x_1(t), x_2(t), \dots, x_n(t) ).
\end{equation}

For each agent $i$, the update rule depends on the opinion of those agents $j$ whose opinion is within a bound $\epsilon$ -- called the \emph{confidence level}.
In this work, the confidence level is a constant fixed at the beginning of the time evolution.
The confidence level naturally defines another set $I$ -- called the set of $i$\textit{'s} $\epsilon$\textit{-insiders}:
\begin{equation}
I(i, X(t), \epsilon) = \left\{ j \bigg| |x_i(t) - x_j(t)| \le \epsilon \right\}.
\end{equation}
Correspondingly, the set of $i$\textit{'s} $\epsilon$\textit{-outsiders} is defined as the complement of $I$:
\begin{equation}
O(i, X(t), \epsilon) = I(i, X(t), \epsilon)^c = \left\{ j \bigg| |x_i(t) - x_j(t)| > \epsilon \right\}.
\end{equation}
With the set of $i$'s $\epsilon$-insiders we can now define the update rule that defines the HK model:
\begin{equation}
x_i(t+1) = \frac{1}{\left| I(i, X(t), \epsilon) \right|} \sum_{j \in I(i, X(t), \epsilon)} x_j(t),
\label{eq:update-rule}
\end{equation}
where $| \cdot |$ refers to the cardinality of the set.
Note that this update rule preserves the \emph{order} of the initial distribution. 
That is, if the initial profile fulfils $x_1(0) \le x_2(0) \le \cdots \le x_n(0)$, then at any time $t$ we will have 
$x_1(t) \le x_2(t) \le \cdots \le x_n(t)$.
This is due to the strictly attracting nature of the update rule, based on averaging over $\epsilon$-insiders. 
Moreover, previous work has shown that the dynamics for the version of the HK model presented here always reaches a steady state (or termination), i.e. we can always find a $\bar{t}$ such that $X(\bar{t}+n) = X(\bar{t})$ $\forall n \in \mathbb{N}$~\cite{Hegselmann-Krause:2002, Blondel:2010, Bhattacharyya:2013, Wedin:2015} 

Based on the definitions above, we can further define a \textit{bounded-confidence process} (BC-process) as a sequence of opinion profiles that are generated by the update rule \eqref{eq:update-rule}.
A BC-process is fully and uniquely determined by the initial opinion distribution $X(0)$ and the chosen confidence level $\epsilon$, since this pair fully characterizes the set of $i$'s $\epsilon$-insiders at every time step and thus fully generates the dynamics in a deterministic fashion.
We therefore refer to the BC-process simply by the ordered pair $\left< X(0), \epsilon \right>$. 

Since a BC-process is fully characterized by the ordered pair $\left< X(0), \epsilon \right>$, there are two main questions we can ask to fully map out the entire space of BC-processes:
\begin{enumerate}
\item Given $X(0)$, find the set of all $\epsilon$ that lead to different BC-processes.
\item Given $\epsilon$, find the initial distributions $X(0)$ that lead to different processes.
\end{enumerate}
The first question is simpler to address and it will be answered in this work.
The second question is much more cumbersome (and in its current formulation, rather ill-defined), as the space of initial states, being defined in a subset of $[0,1]^n$, is uncountably infinite.
It would require the introduction of equivalence classes or some other procedure to work with \emph{collections} of initial states that share certain attributes.
Therefore, we shall not fully address it here, but only briefly comment on it at the end of this paper.

To address the first question, though, Ref.~\cite{Hegselmann:2023} cleverly introduced the concept of $\epsilon$-switches.
An $\epsilon$-switch occurs when the dynamics of the BC-process characterized by the initial state $X(0)$ suddenly changes as $\epsilon$ is progressively increased. 
The change in the dynamics can occur at any time step $t$.
The very first (i.e. smallest) $\epsilon$-switch occurs for a value $\epsilon_1^*$ that equals the smallest distance between two agents with neighboring opinions in the distribution $X(0)$. 
Below this threshold, the dynamics obtained from the initial distribution is trivial, since every agent has no $\epsilon$-insiders. 
The next and nearest larger $\epsilon^*$ that changes the dynamics is the one equal to the distance to the nearest $\epsilon$-outsider that we can find in the \emph{entire} process $\left< X(0), \epsilon_1^* \right>$.
More formally, we define the process of listing up all the successive $\epsilon$-switches as the recursion
\begin{equation}
\epsilon_{j+1}^* = \delta_{\text{min}}^{\text{out}}(X(0), \epsilon_j^*),
\label{eq:delta-recursion}
\end{equation}
where $\delta_{\text{min}}^{\text{out}}(X(0), \epsilon)$ is the minimum element in the set of \emph{all} distances to $\epsilon$-outsiders of \emph{all} agents over \emph{all} time steps $t$ for the BC-process $\left< X(0), \epsilon \right>$, i.e.
\begin{equation}
\delta_{\text{min}}^{\text{out}}(X(0), \epsilon) = \min \left\{ |x_i(t) - x_j(t)| \bigg| t \in \{0, 1, \dots, \bar{t}\}, i \in \mathcal{A}, j \in O(i, X(t), \epsilon) \right\} 
\label{eq:delta-min}
\end{equation}

The recursive evaluation of \eqref{eq:delta-recursion} allows to find all $\epsilon$-switches and thereby completely determines the dynamics for all possible pairs $\left< X(0), \epsilon \right>$, i.e. enumerate all possible BC-processes.
Note that this process yields the set of all $\epsilon$-switches for a given $X(0)$ -- denoted here as $\mathcal{S}(X(0))$ -- that only depends on the structure of the initial opinion distribution $X(0)$.
However, a priori, the cardinality of this set is unknown, which is the subject of one of the conjectures below. 

The procedure for obtaining all $\epsilon$-switches for a few regular distributions was performed systematically in Ref.~\cite{Hegselmann:2023} up to $n=50$, whereby many interesting and previously sidestepped phenomena were observed.
Most notably, a plethora of non-monotonicities in the dynamics and in the number of final opinion clusters was revealed as $\epsilon$ is monotonically increased, which highlights the fundamentally irregular and chaotic nature of the HK model. 

\section{Three conjectures}
Despite the non-monotonic nature of the HK dynamics, strong numerical evidence in Ref.~\cite{Hegselmann:2023} also points towards some stability and regularity in the structure of the HK update rule and the $\epsilon$-switches.
We rephrase here the three main conjectures stated in Ref.~\cite{Hegselmann:2023}.
The goal of this work is to give a formal proof of each of these three conjectures.

\begin{enumerate}
\item The cardinality of the set of all $\epsilon$-switches is finite: $| \mathcal{S}(X(0))| < \infty$, $\forall X(0)$. In other words, there is always a finite number of $\epsilon$-switches for any initial distribution $X(0)$. As a result, the algorithm used to find all $\epsilon$-switches always terminates in a finite time. 
\item If an $\epsilon$-switch $\epsilon^*$ is found at time step $t^*$ during the dynamics of the BC-process $\left< X(0), \epsilon \right>$, then the dynamics of $\left< X(0), \epsilon^* \right>$ will be identical to that of $\left<X(0), \epsilon \right>$ for all times $t \le t^*$. 
\item All fundamental characteristics of a BC-process $\left<X(0), \epsilon \right>$ remain unchanged under positive-affine transformations of the initial distribution, i.e.
\begin{equation}
X(0) \mapsto \tilde{X}(0) \equiv \alpha X(0) + \beta,
\end{equation}
with $\alpha > 0$, $\beta \in \mathbb{R}$, provided that $\epsilon$ is renormalized accordingly, 
\begin{equation}
\epsilon \mapsto \tilde{\epsilon} \equiv \alpha \epsilon.
\end{equation}
These characteristics include the number of final clusters, the time needed to reach the steady state, and the total number of $\epsilon$-switches.
The $\epsilon$-switches are also renormalized in the same way, i.e.
\begin{equation}
\epsilon^* \mapsto \tilde{\epsilon}^* \equiv \alpha \epsilon^*.
\end{equation}
\end{enumerate}

\section{Proofs}
In this section, we prove one by one all the conjectures stated above.

\subsection{Proof of the first conjecture}

First, we remark that the number of $\epsilon$-switches for a given BC-process can either be finite or countably infinite, since it is derived by applying the minimization procedure \eqref{eq:delta-min} over a discrete set of tuples $(i, j, t)$.
The set of all time steps $t$ is finite because -- as mentioned earlier -- $\forall \epsilon, X(0) \: \exists \bar{t}$ such that $X(\bar{t}+1) = X(\bar{t})$, i.e. the dynamics always terminates.
The set $\mathcal{A}$ of all agents $i$ is finite by definition.
The set of all $i$'s $\epsilon$-outsiders is also finite, since it must be a subset of $\mathcal{A}$.
As a result, the set of candidate distances $|x_i(t) - x_j(t)|$ is finite because $t \le \bar{t}$ is finite and $i, j \in \mathcal{A}$ are finite.

The total number of $\epsilon$-switches, i.e. the cardinality $|\mathcal{S}(X(0))|$, is obtained by counting the elements generated by the recursion \eqref{eq:delta-recursion}, which involves searching for a minimum in a discrete and finite set at every step.
Now, since the minimum is searched over a finite number of options for each $\epsilon_j^*$, it will terminate in a finite time.
Obviously, being a minimum, it will produce one element per iteration -- even if there might be multiple (degenerate) opinions with the same $\epsilon$-distance.
Thus, the question is whether the recursion itself terminates or not.

Now consider that
\begin{equation}
| O(i, X(t), \epsilon_2)| < | O(i, X(t), \epsilon_1)| \quad \mathrm{for} \: \epsilon_2 > \epsilon_1.
\end{equation}
This is a consequence or manifestation of the monotonically decreasing total span of the opinions during the dynamics, which in turn is the consequence of the averaging procedure in the HK update rule.

This, together with the fact that the dynamics for any pair $\left< X(0), \epsilon \right>$ always terminates, implies that the sequence of all $\epsilon$-switches -- i.e. $\epsilon_1^*$,  $\epsilon_2^*$,  $\epsilon_3^*$,  $\dots$ -- is a discrete, \emph{strictly increasing} sequence.
Moreover, this sequence has an upper bound ($\epsilon = 1$).
Any discrete, strictly increasing sequence over a finite, bounded set must be finite.
Hence, $| \mathcal{S}(X(0))| < \infty$, and the algorithm that finds all $\epsilon$-switches must terminate in a finite time.
\qed

\subsection{Proof of the second conjecture}

If an $\epsilon$-switch $\epsilon^*$ is found at $t^*$ for $\left< X(0), \epsilon \right>$, it means that 
\begin{equation}
t^* = \mathrm{argmin}_t \left\{ |x_i(t) - x_j(t) | \bigg| i \in \mathcal{A}, t \in \{0, 1, \dots, \bar{t} \}, j \in O(i, X(t), \epsilon) \right\}.
\end{equation}
If there are more than one such times $t^*$, let us take the smallest one.
We then have
\begin{equation}
 |x_i(t) - x_j(t)|  > |x_i(t^*) - x_j(t^*)| = \epsilon^*, \qquad \forall t < t^*, \forall i, \forall j  \in O(i, X(t), \epsilon)
\end{equation}
due to the minimum condition and
\begin{equation}
|x_i(t) - x_j(t)| \le \epsilon, \qquad \forall t < t^*, \forall i, \forall j \in I(i, X(t), \epsilon)
\end{equation}
due to the definition of the insider condition.
All the combinations of distances that occur in the dynamics before $t^*$ are either $i$'s $\epsilon$-insiders and thus $\epsilon^*$-insiders too because $\epsilon^* > \epsilon$, or are larger than the minimum $\epsilon^*$ if they are $i$'s $\epsilon$-outsiders.
In other words, up to the (first) time $t^*$, where some outsider distance attains $\epsilon$, raising the threshold from $\epsilon$ to $\epsilon^*$ neither removes insiders nor adds any new insiders before $t^*$.
Therefore the insider sets (and thus the updates) coincide for all $t \le t^*$ for $\left< X(0), \epsilon \right>$ and $\left< X(0), \epsilon^* \right>$.

By the same token, at time $t^*$, there will be at least one pair $(i,j)$ s.t. $|x_i(t^*) - x_j(t^*)| = \epsilon^*$, which will lead to a different grouping and number of insiders for a particular $i$, which in turn will change the dynamics from that time onwards.
\qed

\begin{figure}
\centering
\includegraphics[scale=0.6]{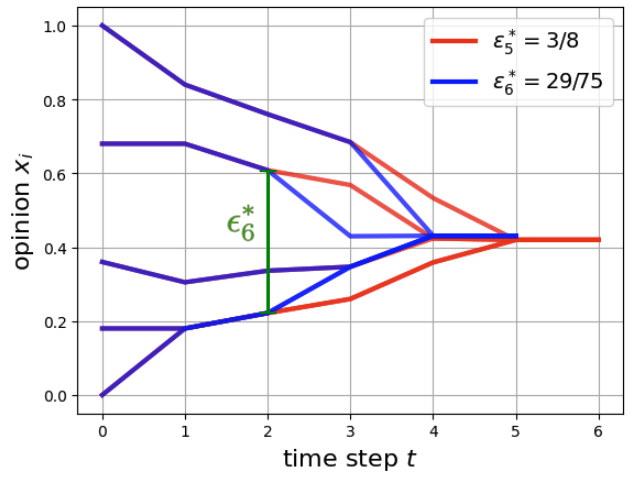}
\caption{The opinion dynamics of the initial distribution $X(0)=[0.0, 0.18, 0.36, 0.68, 1.0]$ for two consecutive $\epsilon$-switches, $\epsilon_5^*=3/8$ and $\epsilon_6^*=29/75$. The $\epsilon$-switch for $\left< X(0), \epsilon_5^* \right>$ occurs at time $t^*=2$, as indicated by the green line.
As stated and proven in the second conjecture, the dynamics up to $t*$ for $\left< X(0), \epsilon_5^* \right>$ and $\left< X(0), \epsilon_6^* \right>$ is identical.}
\label{fig:second-conj}
\end{figure}

An illustration of the second conjecture is presented in Fig.~\ref{fig:second-conj}.

\subsection{Proof of the third conjecture}
Without loss of generality we restrict ourselves to those transformations that keep $X(0) \in [0,1]$.
Otherwise, we can simply extend the range of validity for $\epsilon$ to a larger interval $[\beta, \alpha + \beta]$.
Also note that any positive-affine transformation preserves the order of the agents' opinions in the distribution $X(0)$.

The update rule for the BC-process is given in \eqref{eq:update-rule} and depends crucially on $I(i, X(t), \epsilon)$.
Since $I(i, X(t), \epsilon)$ only depends on the relative distances between agents' opinions, it is unaffected by any overall shift $\beta$.
However, it is affected by the overall (uniform) stretch of the initial distribution, encoded in $\alpha$.
The set is modified to
\begin{align}
\tilde{I}(i, \tilde{X}(t), \epsilon) &= \left\{ j \bigg| | \alpha x_i(t) - \alpha x_j(t) | \le \epsilon \right\} \\
&= \left\{ j \bigg| | x_i(t) -  x_j(t) | \le \epsilon/\alpha \right\}  \\
&= I(i, X(t), \epsilon/\alpha).
\end{align}
Therefore, by rescaling $\epsilon \mapsto \tilde{\epsilon} = \alpha \epsilon$, we obtain
\begin{equation}
\tilde{I}(i, \tilde{X}(t), \tilde{\epsilon}) =  I(i, X(t), \epsilon).
\end{equation}

Now, since the distribution of insiders is what determines the dynamics via the update rule and the stretch factor $\alpha$ doesn't impact it (provided $\epsilon$ is rescaled), the entire time evolution will be the same.
Concretely:
\begin{align}
\tilde{x}_i(t+1) = \frac{ \sum_{j \in \tilde{I}(i, \tilde{X}(t), \tilde{\epsilon})} \tilde{x}_j(t)}{| \tilde{I}(i, \tilde{X}(t), \tilde{\epsilon})|} &\iff \alpha x_i(t+1) + \beta = \frac{1}{| I(i, X(t), \epsilon)|} \sum_{j \in I(i, X(t), \epsilon)} ( \alpha x_j(t) + \beta ) \\
&\iff \alpha x_i(t+1) + \beta = \alpha \frac{\sum_{j \in I(i, X(t), \epsilon)} x_j(t)}{| I(i, X(t), \epsilon)|} + \frac{\sum_{j \in I(i, X(t), \epsilon)} \beta}{| I(i, X(t), \epsilon)|} \\
&\iff \alpha x_i(t+1) + \beta = \alpha \frac{\sum_{j \in I(i, X(t), \epsilon)} x_j(t)}{| I(i, X(t), \epsilon)|} + \beta \\
&\iff \alpha x_i(t+1) = \alpha \frac{\sum_{j \in I(i, X(t), \epsilon)} x_j(t)}{| I(i, X(t), \epsilon)|} \\
&\iff x_i(t+1) = \frac{\sum_{j \in I(i, X(t), \epsilon)} x_j(t)}{| I(i, X(t), \epsilon)|},
\end{align}
i.e. the update rule is the same.

As a corollary, we can see that the $\epsilon$-switches will occur at the values $\tilde{\epsilon}^* = \alpha \epsilon^*$, since the outsider set will undergo the same transformation as the insider set (the only change is a reversed inequality sign) and fulfill
\begin{equation}
\tilde{O}(i, \tilde{X}(t), \tilde{\epsilon}) = O(i, X(t), \epsilon).
\end{equation}
The $\epsilon$-switches are then defined from the minimum, which acquires an overall shift too:
\begin{align}
\tilde{\epsilon}^* &= \min \left\{ | \tilde{x}_i(t) - \tilde{x}_j(t) | \bigg| t \in \{0, 1, \dots, \bar{t} \}, i \in \mathcal{A}, j \in \tilde{O}(i, \tilde{X}(t), \tilde{\epsilon}) \right\} \\
&= \min \left\{ \alpha | x_i(t) - x_j(t) | \bigg| t \in \{0, 1, \dots, \bar{t} \}, i \in \mathcal{A}, j \in O(i, X(t), \epsilon) \right\} \\
&= \alpha \epsilon^*.
\end{align}
Therefore the list of $\epsilon$-switches scales as $\tilde{\epsilon}^*_k = \alpha \epsilon_k^*$ with identical multiplicities and termination time (in steps) unchanged.
\qed

\begin{figure}
\centering
\includegraphics[scale=0.47]{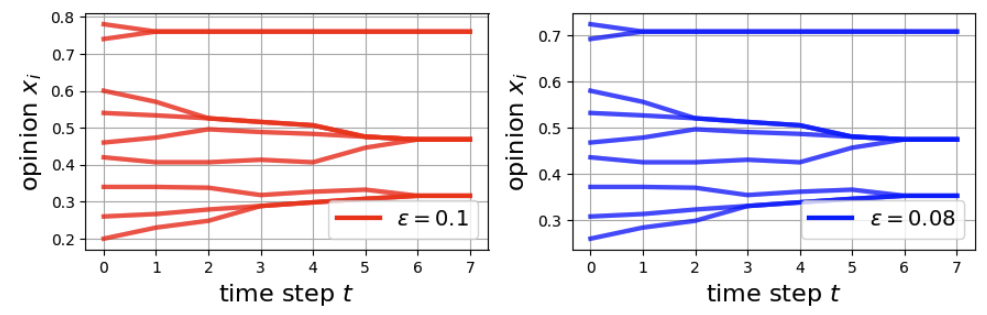}
\caption{Comparison between the BC-process $\left< X(0), \epsilon \right>$ and $\left<\alpha X(0) + \beta, \alpha \epsilon \right>$ for $X(0) = [0.2, 0.26, 0.34, 0.42, 0.46, 0.54, 0.6, 0.74, 0.78]$, $\alpha=0.8$, $\beta=0.1$ and $\epsilon=0.1$.
The dynamics of the two BC-processes is completely equivalent once the $\beta$ shift is taken into account, including the opinion evolution, the steady-state distribution, and the location of $\epsilon$-switches, corroborating the third conjecture.}
\label{fig:third-conj}
\end{figure}

A numerical example illustrating the third conjecture is presented in Fig.~\ref{fig:third-conj}.

\section{Outlook: on the universality of initial distributions}

The results above proved that the essential dynamical features of a BC-process are preserved under positive-affine transformations of the initial opinion profile, irrespective of its distribution. 
This invariance implies that, from the point of view of the HK dynamics, many distinct-looking initial states are in fact dynamically equivalent. 
In practical terms, stretching or shifting the entire configuration merely rescales the relevant $\epsilon$ thresholds and leaves the temporal evolution of the opinion structure unchanged.

This observation naturally invites a more general question: what other transformations on $X(0)$ preserve the main qualitative features of the dynamics? 
Beyond simple affine mappings, one could consider non-linear, order-preserving transformations, or operations that only affect the relative spacing of certain opinion clusters while maintaining key combinatorial properties of the insider-outsider relationships. 
Identifying such transformations would help to reduce the effective dimensionality of the space of initial conditions and to classify BC-processes into broad families with identical or closely related behaviors.

From a mathematical perspective, this suggests introducing a notion of conjugacy or equivalence classes for initial opinion distributions. 
Two distributions $X(0)$ and $Y(0)$ would be considered equivalent if there exists a transformation $T$ -- invertible on the relevant opinion space -- such that $T$ maps the insider-outsider relations of $X(0)$ to those of $Y(0)$ for all time steps. 
The study of these equivalence classes could draw inspiration from functional analysis and dynamical systems theory, where conjugacies are used to classify systems with qualitatively identical trajectories~\cite{Katok:1995, Devaney:2003}.

Future work could aim to systematically construct and characterize these mappings, possibly leading to a universal parametrization of BC-processes. 
Such a framework would not only deepen our theoretical understanding of opinion dynamics, but also simplify computational explorations by focusing on representative initial states from each equivalence class, rather than exhaustively scanning the entire high-dimensional space of initial conditions.

\end{document}